\documentstyle[floats,aps]{revtex}
\begin{document}
\input epsf

\hsize=6.5truein
\hoffset=0.0truein
\vsize=9.0truein
\voffset=0.2truein
\hfuzz=0.1pt
\vfuzz=0.1pt
\parskip=\medskipamount
\overfullrule=0pt

\font\twelvemib=cmmib10 scaled 1200
\font\elevenmib=cmmib10 scaled 1095
\font\tenmib=cmmib10
\font\eightmib=cmmib10 scaled 800
\font\sixmib=cmmib10 scaled 667
\skewchar\elevenmib='177
\newfam\mibfam
\def\mib{\fam\mibfam\tenmib}
\textfont\mibfam=\tenmib
\scriptfont\mibfam=\eightmib
\scriptscriptfont\mibfam=\sixmib

\draft

\twocolumn[\hsize\textwidth\columnwidth\hsize\csname  
@twocolumnfalse\endcsname
\title{Collapse of a Bose Condensate with Attractive Interactions}

\author{Jos{\'e} A. Freire$^1$ and Daniel P. Arovas$^2$}

\address{
${}^1$Centro Brasileiro de Pequisas F{\'\i}sicas,
Rua Dr. Xavier Sigaud 150, Rio de Janeiro-RJ,
22290-180 BRAZIL\\
${}^2$Department of Physics, University of California at San Diego,
La Jolla, CA 92093, USA}

\date{\today}

\maketitle

\begin{abstract}
We examine the Gross-Pitaevskii (GP) model for Bose-Einstein condensates in
parabolic traps with attractive interactions.  The decay of metastable
condensates is investigated by applying the instanton formalism to the GP
field theory.  Employing various dynamical trial states, we derive within
a coherent state path integral approach a collective coordinate description
in terms of the condensate radius, in agreement with (and extending)
earlier results.  We then solve numerically for the complete instanton field
configuration and compare with the collective coordinate approach.
Adjusting only the effective mass of the collective coordinate, the two
schemes are then in good agreement.
\end{abstract}

\pacs{PACS numbers:}
\vskip2pc]

\narrowtext

\section{Introduction}
It is well known that a Bose gas with purely attractive interactions is
unstable and will collapse to a state with no thermodynamic limit.
However, in the presence of a harmonic confining potential, Ruprecht
{\it et al.\/}\ \cite{rhb95} have shown that a {\it metastable\/}
condensate exists, provided the particle number $N$ is sufficiently small.
The condition for a metastable condensate is $N<N_ {\rm c}$, with
$N_ {\rm c}\propto \ell/|a|$,
where $\ell$ is the quantum confinement length associated with the trap,
and $a$ is the $s$-wave scattering length; $a<0$ corresponds to an attractive
$s$-wave pseudopotential.  Such is the situation with ${}^7$Li, for which
$a=-(14.5\pm 0.4)\,a_{\scriptscriptstyle\rm B}$,
where $a_{\scriptscriptstyle\rm B}$ is the Bohr radius
\cite{ams95}.  Indeed, experiments by Bradley
{\it et al.\/}\ \cite{bsh97} demonstrated
the existence of metastable ${}^7$Li condensates of limited particle number
($N<N_ {\rm c}$).  Experiment and theory are in
rough agreement on the value of $N_ {\rm c}$.

The metastable condensate has a finite lifetime -- it tunnels quantum
mechanically to an unstable higher density state, which then collapses.
The tunneling process was described by Stoof \cite{sto97} in terms of a
collective coordinate $q(t)$ which parameterizes a Gaussian condensate density,
\begin{equation}
\rho( {\mib r};q(t))=N\, \left ( {1\over\pi q^2(t)} \right ) ^{3/2}\,
\exp \left ( -{ {\mib r}^2\over q^2(t)} \right ) \ .
\label{gauss}
\end{equation}

The quantum mechanics of this collective coordinate arises from the
more microscopic Gross-Pitaevskii (GP) field theory \cite{gp61},
which is itself a simplification inasmuch as interatomic potentials
are therein replaced by an effective $s$-wave pseudopotential
$g\,\delta( {\mib r})$.  To compute the tunneling rate, one must examine
the Euclidean action $ S ^{\vphantom{\dagger}}_ {\rm E}$ for the GP model,
\begin{eqnarray}
S ^{\vphantom{\dagger}}_ {\rm E}/\hbar&=&
\int\limits_{- \frac{1}{2}\beta}^{ \frac{1}{2}\beta}\!\! d\tau \!\!\int\!\!
d^3\!r\, \left\{ {\bar\psi}\, {\partial}_\tau\psi
+ \frac{1}{2}\, {\vec\nabla} {\bar\psi}\cdot
{\vec\nabla}\psi\vphantom{\sum}\right.\nonumber\\
&&\qquad\qquad\left.\vphantom{\sum}+ \frac{1}{2}\, r^2\, {\bar\psi}\psi+
\frac{1}{2}\, g\, ( {\bar\psi}\psi)^2\right\},
\label{SE}
\end{eqnarray}
\begin{figure} [!t]
\centering
\leavevmode
\epsfxsize=8cm
\epsfysize=8cm
\epsfbox[18 144 592 718] {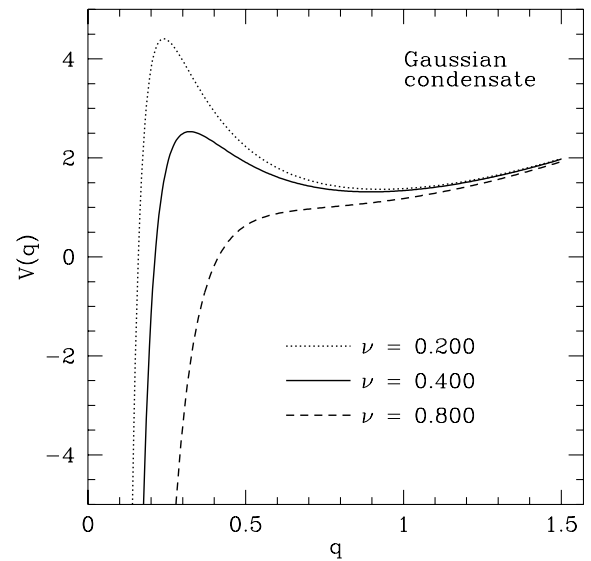}
\caption[]
{\label{fig1} The potential energy $V(q)$ derived from the Gaussian
condensate profile of equation \ref{gauss} for three values of $\nu$.
When $\nu<\nu_ {\rm c}=16/15^{5/4}\simeq 0.542$ a metastable condensate
exists.}
\end{figure}
Here and below distance is measured in
units of $\ell=\sqrt{\hbar/m\omega}$, where $m$ is the boson mass and $\omega$
is the trap frequency, time in units of $\omega^{-1}$, and the fields
$\psi$ and $ {\bar\psi}$ in units of $\ell^{-3/2}$.  The coupling
constant is then $g=4\pi a/\ell$, where $a$ is the (negative) $s$-wave
scattering length.  Periodic boundary conditions are enforced over
the imaginary time domain $\tau\in[- \frac{1}{2}\beta, \frac{1}{2}\beta]$ where
$\beta$ is the inverse temperature (units of $\hbar\omega$).
Writing the complex order parameter field
$\psi=\sqrt{\rho}\,\exp(i\theta)$ in terms of its amplitude and phase,
Stoof formally integrates out the phase
field $\theta( {\mib r},\tau)$, resulting
in an effective action $S^{\vphantom{\dagger}}_{\rm E}
[\rho( {\mib r},\tau)]$ in terms of the density alone.
Using the trial density from equation \ref{gauss}, the action functional
becomes
\begin{equation}
S^{\vphantom{\dagger}}_{\rm E}/\hbar=
N\!\!\int\!d\tau\,\left\{ \frac{1}{2} m^*  {\dot q}^2 + V(q)\right\}\ ,
\label{Seff}
\end{equation}
where $m^*$ is an effective mass and
\begin{equation}
V(q)={3\over 4}\,{1\over q^2}+{3\over 4}\,q^2-{\nu\over\sqrt{2\pi}}\,
{1\over q^3}
\label{Vgau}
\end{equation}
and $\nu=N |a|/\ell=N|g|/4\pi$.  A large value of $N$ justifies the
semiclassical treatment of this problem.  The above effective potential $V(q)$
reflects the kinetic energy of confinement, the trap potential, and the
attractive interparticle potential, respectively.
Provided $\nu<\nu_{\rm c}=16/15^{5/4}$,
$V(q)$ has a local minimum at $q=q^{\vphantom{\dagger}}_0$,
corresponding to a metastable
condensate.  This state is isoenergetic with a denser condensate at
$q=q^{\vphantom{\dagger}}_{\rm n}< q^{\vphantom{\dagger}}_0$,
and by considering motion in the inverted potential
$-V(q)$ one constructs the usual bounce trajectory
$q^{\vphantom{\dagger}}_ {\rm b}(\tau)$ \cite{col85}
and from it the tunneling rate
$\Gamma\propto\exp(-\Delta S ^{\vphantom{\dagger}}_{\rm E}/\hbar)$, where
\begin{eqnarray}
\Delta  S ^{\vphantom{\dagger}}_ {\rm E}&=&  S ^{\vphantom{\dagger}}_{\rm E}
[q^{\vphantom{\dagger}}_{\rm b}(\tau)]-S^{\vphantom{\dagger}}_{\rm E}
[q^{\vphantom{\dagger}}_0]\nonumber\\
&=&N\hbar\sqrt{8m^*}\int_{q^{\vphantom{\dagger}}_{\rm n}
}^{q^{\vphantom{\dagger}}_0}\!\!\!dq\,
\sqrt{V(q)-V(q^{\vphantom{\dagger}}_0)}\ .
\label{dS}
\end{eqnarray}

In this paper we shall use the formalism of coherent state path integration
\cite{no88} to treat the problem of the collapsing metastable condensate.
The instanton configuration for the GP field theory of equation \ref{SE}
is described by two complex {\it fields\/} $\psi( {\mib r},\tau)$ and
$ {\bar\psi}$.  Restricting our attention to trial field configurations,
we recover the collective coordinate effective action of equation \ref{Seff}.

\section{Dynamics and Quantum Tunneling: Collective Coordinate Approach}

Our goal is to numerically solve for the instanton of the GP action
in the zero temperature limit by numerically solving for the saddle
point of the action-extremizing equations.  First, though, we 
concern ourselves with trial functions $\psi( {\mib r},\tau)$
which yield a one-parameter effective action in the form of equation
\ref{Seff}.  In the zero temperature $(\beta\to\infty)$ limit, the
initial and final configurations, $\psi(\pm \frac{1}{2}\beta, {\mib r})$ should 
represent a metastable condensate.  At the temporal
midpoint $\tau=0$, $\psi(0, {\mib r})$ represents the state of the field after
it emerges from the tunnel barrier,  {\it i.e.\/}\ a denser
condensate whose real time
evolution (according to the nonlinear Schr{\"o}dinger equation) is unstable
towards collapse.  To interpolate between these states, let us make the
dynamical {\it Ansatz}
\begin{eqnarray}
\psi( {\mib r},\tau)&=&A\,\exp(- \alpha\, r^2)
\label{tri1}\\
 {\bar\psi}&=& {\bar A}\,\exp(- {\bar\alpha}\, r^2)\ .
\label{tri2}
\end{eqnarray}
where $A$, $ {\bar A}$, $\alpha$, and ${\bar\alpha}$ are functions of the time.
Number conservation relates these quantities, since
\begin{equation}
N=\int d^3\!r\, {\bar\psi}\psi = {\pi^{3/2}\,A
{\bar A}\over ( \alpha+ {\bar\alpha})^{3/2}}\ .
\end{equation}
Inserting our {\it Ansatz\/} into equation \ref{SE}, we find
\begin{eqnarray}
S^{\vphantom{\dagger}}_ {\rm E}/\hbar&=&N\int\!d\tau
\left\{-{3\over 2}\,{{\dot\alpha}\over\alpha+ {\bar\alpha}}
+3\,{\alpha {\bar\alpha}\over  \alpha+ {\bar\alpha}}\right.\nonumber\\
&&\qquad\qquad\left.+{3\over 4}\,{1\over  \alpha+ {\bar\alpha}}
-{\nu\over\sqrt{2\pi}}\,( \alpha+ {\bar\alpha})^{3/2}\right\}\ ,
\end{eqnarray}
plus a term $N\!\int\!d\tau\, {\partial}_\tau\ln A$ which vanishes owing to
periodicity.

\begin{figure} [!t]
\centering
\leavevmode
\epsfxsize=8cm
\epsfysize=10cm
\epsfbox[18 144 592 718] {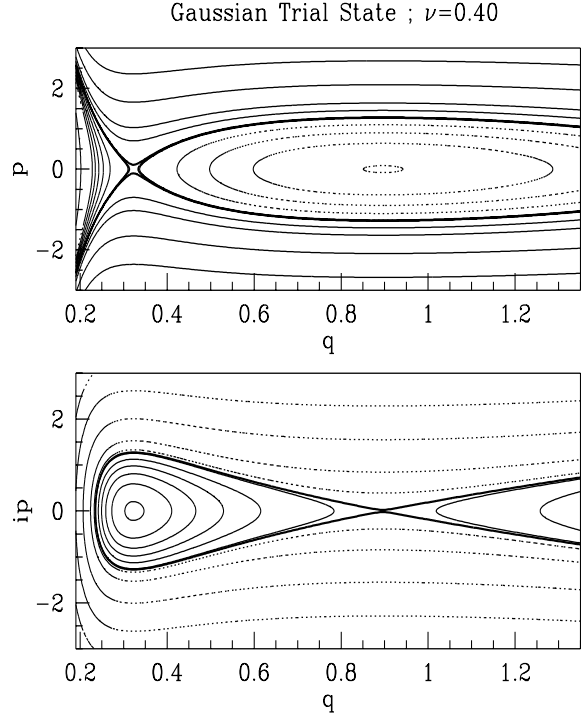}
\caption[]
{\label{fig2} Contour plots of $H(p,q)$ (top) and $H(ip,q)$ (bottom)
for the Gaussian trial states ($\nu=0.40$). Solid lines represent highs and
dotted lines lows.  (In the top graph, the low contours increase linearly
while the high ones increase quadratically.  In the bottom graph,
the situation is reversed.)}
\end{figure}
We emphasize that during the instanton event, $ \alpha$ and $ {\bar\alpha}$ are
{\it not\/} related by complex conjugation.  This feature of the instanton
solution is familiar, for if one treats the problem of quantum tunneling
in a potential $V(x)$ using coherent states, the saddle point path is
determined by the equations
\begin{eqnarray}
{{\partial} z/ {\partial}\tau}&=&-{ {\partial} H/ {\partial} {\bar z}}\\
{{\partial} {\bar z}/ {\partial}\tau}&=&+{ {\partial} H/ {\partial} z}\ ,
\end{eqnarray}
where $z=(x+ip)/\sqrt{2}$, $ {\bar z}=(x-ip)/\sqrt{2}$,
and $H= \frac{1}{2} p^2 + V(x)$.
During the instanton event, $ {\bar z}\ne z^*$ -- indeed
$ {\bar z}=z^*(-\tau)$ -- so the momentum $p(\tau)$ is imaginary.

\begin{figure} [!t]
\centering
\leavevmode
\epsfxsize=8cm
\epsfysize=8cm
\epsfbox[18 144 592 718] {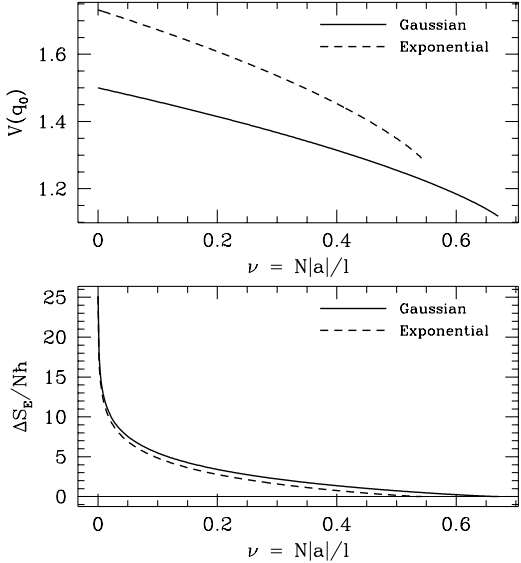}
\caption[]
{\label{fig3} Numerically obtained results for the energy
$V(q^{\vphantom{\dagger}}_0)$ of the metastable state and action
$\Delta S^{\vphantom{\dagger}}_{\rm E}/N\hbar$ as a function of $\nu$.
Results are shown for the Gaussian and the Exponential
trial condensate functions.}
\end{figure}

In fact, we can change variables to $(q,p)$, where
\begin{eqnarray}
\alpha&\equiv& \frac{1}{2}(q^2+iq p^{-1})\\
{\bar\alpha}&\equiv& \frac{1}{2}(q^2-iq p^{-1})
\end{eqnarray}
so that
\begin{eqnarray}
S^{\vphantom{\dagger}}_{\rm E}/\hbar&=&N\int\!d\tau
\left\{{3\over 4}\,i\,(p{\dot q}-q{\dot p})
+{3\over 4}\,(p^2+q^{-2})\right.\nonumber\\
&&\qquad\left.+{3\over 4}\,q^2-{\nu\over\sqrt{2\pi}}\,q^{-3}\right\}\ .
\end{eqnarray}
Let us define the Hamiltonian
\begin{equation}
H(p,q)=\frac{3}{4}\, p^2 + \frac{3}{4}\, q^{-2}+\frac{3}{4}\,q^2
-{\nu\over\sqrt{2\pi}}\,q^{-3}\ .
\end{equation}
The equations of motion are then
\begin{eqnarray}
-\frac{3}{2}\,i\,{\dot q}&=& {\partial} H/ {\partial} p\\
\frac{3}{2}\,i\,{\dot p}&=& {\partial} H/ {\partial} q\ ,
\end{eqnarray}
from which we see that $ip$ is real during the instanton event, as usual.

In figure \ref{fig2} we show contour plots of $H(p,q)$ and $H(ip,q)$ for
$\nu=0.40$.  The minimum of $H(p,q)$ occurs at
$(0,q^{\vphantom{\dagger}}_0)$ while $q^{\vphantom{\dagger}}_1$
is the location of the maximum of $V(q)$.
The saddle point is traversed during the instanton by $p$ becoming imaginary
and $ip$ moving around the contour line
$H(ip,q)=V(q^{\vphantom{\dagger}}_0)$, which encircles
the local maximum $(0, q^{\vphantom{\dagger}}_1)$
of $H(ip,q)$.  At $\tau=0$, $\psi$ describes
the nucleated state; this is of course denser,
since $q(\tau=0)=q^{\vphantom{\dagger}}_{\rm n} <  q^{\vphantom{\dagger}}_0$.

Since the dynamics preserves $H(ip,q)$, the instanton action is given by
\begin{equation}
S^{\vphantom{\dagger}}_{\rm E}/N\hbar=
\beta V(q^{\vphantom{\dagger}}_0)+\frac{3}{2}i\hskip-0.28truein
\oint\limits_{H(ip,q)=V( q ^{\vphantom{\dagger}}_0)}\hskip-0.28truein p\,dq\ .
\end{equation}
The difference $\Delta S ^{\vphantom{\dagger}}_{\rm E}/N\hbar=
S ^{\vphantom{\dagger}}_{\rm E}/N\hbar-\beta V(q^{\vphantom{\dagger}}_0)$
is proportional to the area enclosed by the contour.

Integrating out the momentum $p(\tau)$, we obtain the action of
equations \ref{Seff} and \ref{Vgau}, with $m^*=\frac{3}{2}$.  Stoof
\cite{sto97} states that this is in fact the exact value
for Gaussian trial states.  Within his approximation scheme, he obtains
$m^*\simeq 0.27$.

Similar results are obtained using exponential trial functions\cite{foot1}
\begin{eqnarray}
\psi( {\mib r},\tau)&=&A\,\exp(- \alpha\, r)
\label{tri3}\\
 {\bar\psi}&=& {\bar A}\,\exp(- {\bar\alpha}\, r)\ ,
\label{tri4}
\end{eqnarray}
from which one derives
\begin{eqnarray}
S^{\vphantom{\dagger}}_{\rm E}/\hbar&=&N\int\!d\tau
\left\{-{3\,{\dot\alpha}\over  \alpha+ {\bar\alpha}}
+ \frac{1}{2}\, \alpha {\bar\alpha}+6\,
( \alpha+ {\bar\alpha})^{-2}\right.\nonumber\\
&&\qquad\left.-{\nu\over 32}\,( \alpha+ {\bar\alpha})^3
\right\}\ .
\end{eqnarray}
With $ \alpha\equiv \frac{1}{2}$ and $ {\bar\alpha}\equiv \frac{1}{2}$,
one obtains results similar to those for the Gaussian model.
Integrating out $p$ again generates equation \ref{Seff} but with
$m^*=9$ and
\begin{equation}
V(q)={1\over 8}\,{1\over q^2}+6\,q^2-{\nu\over 32}\,{1\over q^3}\ .
\end{equation}

Numerical results (see figure \ref{fig3}) show that for any given value of
$\nu$, that the metastable state energy
$V( q ^{\vphantom{\dagger}}_0)$ is lower
for the Gaussian model.  The computed action
$\Delta  S ^{\vphantom{\dagger}}_{\rm E}/\hbar$ differs little, however
(although the maximum value $\nu^{\vphantom{\dagger}}_{\rm max}$
is model-dependent).

If $q$ is the condensate radius, the functional form
\begin{equation}
V(q)=b\,q^{-2} + c\,q^2 -d\,\nu\, q^{-3}
\end{equation}
is generic.  Setting $ {\partial} V/ {\partial} q=0$ gives
$q= q ^{\vphantom{\dagger}}_0$ through
\begin{equation}
\nu={2\over 3\,d}\,q\,(b - c\,q^4)\equiv f(q)
\end{equation}
which is positive for $0 < q <  Q\equiv {\root 4 \of {b/c}}$.
Stable condensates have $q> q ^{\vphantom{\dagger}}_{\rm c}$
where $f'( q ^{\vphantom{\dagger}}_{\rm c})=0$; this gives
$ q ^{\vphantom{\dagger}}_{\rm c}=Q/{\root 4 \of 5}={\root 4 \of {b/5c}}$.
Thus $\nu$ is restricted to lie in the range
$[0, \nu ^{\vphantom{\dagger}}_{\rm c}]$, where
\begin{equation}
\nu^{\vphantom{\dagger}}_{\rm c}=f(q^{\vphantom{\dagger}}_{\rm c})=
{8\,b\over 15\, d} \left ( {b\over 5\, c} \right ) ^{1/4}\ .
\end{equation}
Results for the exponential and Gaussian trial condensates
are summarized in the following table:

\begin{center}
\begin{tabular}{c||c|c|c|c|c|}
&  \hskip 0.2cm condensate  \hskip 0.2cm &  \hskip 0.2cm $b$  \hskip 0.2cm  &
\hskip 0.2cm $c$  \hskip 0.2cm &  \hskip 0.2cm $d$  \hskip 0.2cm &
\hskip 0.2cm $m^*$  \hskip 0.2cm\\
\hline\hline
&&&&&\\
Gaussian \hskip 0.2cm &  \hskip 0.2cm $\psi=A\exp(-\alpha\, r^2)$
\hskip 0.2cm &  \hskip 0.2cm $\frac{3}{4}$  \hskip 0.2cm &
\hskip 0.2cm $\frac{3}{4}$  \hskip 0.2cm &  \hskip 0.2cm
$\frac{1}{\sqrt{2\pi}}$  \hskip 0.2cm &  \hskip 0.2cm$\frac{3}{2}$
\hskip 0.2cm\\
&&&&&\\
\hline
&&&&&\\
Exponential \hskip 0.2cm &  \hskip 0.2cm $\psi=A\exp(-\alpha\, r)$
\hskip 0.2cm &  \hskip 0.2cm $\frac{1}{8}$  \hskip 0.2cm &
\hskip 0.2cm$6$  \hskip 0.2cm &  \hskip 0.2cm $\frac{1}{32}$
\hskip 0.2cm &  \hskip 0.2cm $9$  \hskip 0.2cm \\
&&&&&\\
\hline
\end{tabular}
\end{center}

\begin{figure} [!t]
\centering
\leavevmode
\epsfxsize=8cm
\epsfysize=8cm
\epsfbox[18 144 592 718] {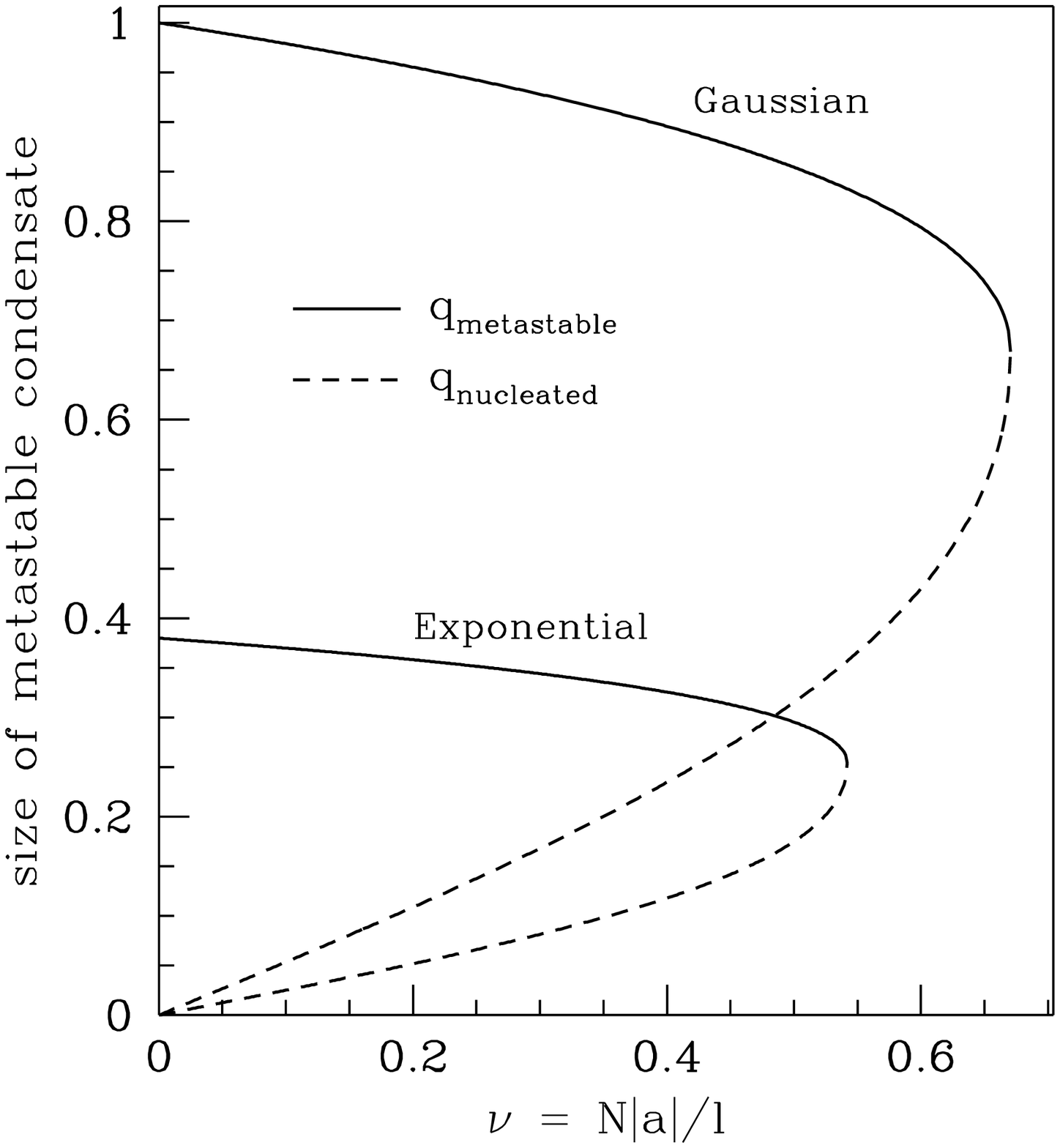}
\caption[]
{\label{fig4} Trial state results for $q^{\vphantom{\dagger}}_0(\nu)$ (solid)
and $ q ^{\vphantom{\dagger}}_ {\rm n}(\nu)$ (dashed).}
\end{figure}
The chemical potential is given by
\begin{equation}
\mu=-{1\over 3}\,b\, q_0^{-2} + {7\over 3}\,c\, q_0^2\ ,
\end{equation}
hence we find
\begin{eqnarray}
\nu=0&\colon\ & q={\root 4 \of {b/c}}\ ,\  \mu=2\,\sqrt{bc}\\
\nu= \nu ^{\vphantom{\dagger}}_{\rm c}&
\colon\ & q={\root 4 \of {b/5c}}\ ,
\  \mu=\mu^{\vphantom{\dagger}}_{\rm c}\equiv{2\over 3\sqrt{5}}\,\sqrt{bc}\ .
\end{eqnarray}
Thus, in the vicinity of the critical point $q=q^{\vphantom{\dagger}}_{\rm c}$,
one has
\begin{equation}
{\nu\over\nu^{\vphantom{\dagger}}_{\rm c}}=1-{5\over 288}\left(
{\mu\over\mu^{\vphantom{\dagger}}_{\rm c}}-1\right)^2 + \ldots
\end{equation}

\begin{figure} [!t]
\centering
\leavevmode
\epsfxsize=8cm
\epsfysize=8cm
\epsfbox[18 144 592 718] {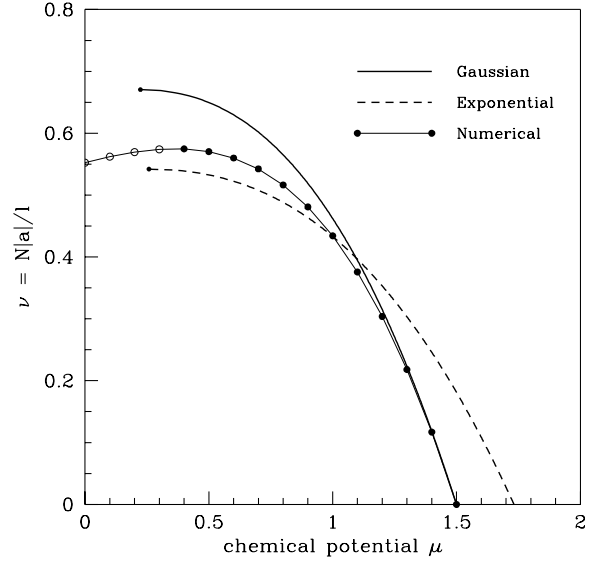}
\caption[]
{\label{fig5} Scaled particle number {\it versus\/} chemical potential
for both Gaussian and exponential models.
For $\mu<\mu_{\rm c}$, the condensate
is unstable; the curves terminate in a dot at these points.  The numerical
results obtained by solving the Bogoliubov equations are shown for comparison.
Open circles denote unstable condensates.}
\end{figure}

\section{Beyond the Collective Coordinate Approach}

We now discuss a full solution of the GP equations, where all degrees of
the condensate are retained.
We begin with the
condensate itself and the real time Bogoliubov equations for excited states.
We begin with the nonlinear Schr\"odinger equation for $\phi\equiv\sqrt{|g|}\,
\psi$,
\begin{equation}
i {\partial} ^{\vphantom{\dagger}}_t\phi=
-\frac{1}{2}\nabla^2\phi+ \frac{1}{2} r^2\phi
- ( {\bar\phi}\phi)\phi -\mu\phi\ ,
\label{tdnlse}
\end{equation} 
where $\mu$ is the dimensionless chemical potential in units of $\hbar\omega$
(we now work in the grand canonical ensemble).
We are interested in stationary states, and we consider only real, radially
symmetric solutions, for which we write $\phi(r)=R(r)/r$.  The radial
wave equation for $R(r)$ is
\begin{equation}
- \frac{1}{2}{d^2\!R\over dr^2}+( \frac{1}{2} r^2-\mu)R -{R^3\over r^2}=0
\label{Reqn}
\end{equation} 
subject to the boundary conditions $R(0)=R(\infty)=0$.
We solved this equation on a one-dimensional grid and used a relaxation
method \cite{numrec}.  The size of the system was chosen large enough to
capture the structure of the wave function.   We used as many as $1,000$
grid points in the range $r \in [0,10]$.  The results were checked for
independence on the system size.  This was done for various positive values
of the chemical potential $\mu$; we obtained the ground and several excited
states.  The number of atoms is given by $N=4\pi\nu/|g|$, with
\begin{equation}
\nu=\int_0^\infty\!\!\!dr\,R^2(r)\ .
\end{equation}

The dependence of $\nu$ on the chemical potential is shown in figure
\ref{fig5} along with the predictions of the trial condensate models.
Stationary condensates exist for all values of $\mu < 1.5$.  The
stability of these states can only be checked by a numerical integration of
the time dependent nonlinear Schr{\"o}dinger equation (NLSE) \cite{rhb95}, or,
as we did, by computing the spectrum of fluctuations above these condensates.

\begin{figure}[!t]
\centering
\leavevmode
\epsfxsize=8cm
\epsfysize=5cm
\epsfbox[18 144 592 718] {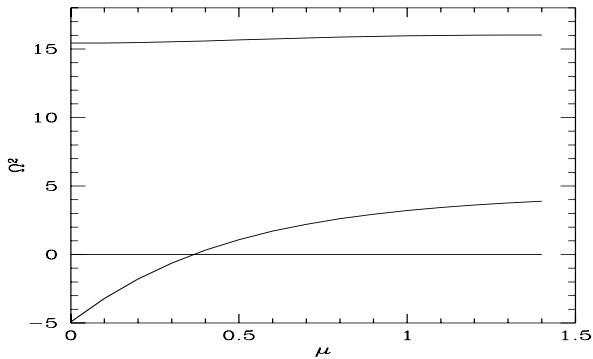}
\caption[]
{\label{fig6} The three lowest values of $\Omega^2$, eigenfrequencies
of the Bogoliubov equations, as a function of the chemical potential.
A negative $\Omega^2$ indicates an unstable condensate.}
\end{figure}

Once the ground state solution $R_0$ is found, we solve for the
spectrum of fluctuations, writing
$\phi= \phi ^{\vphantom{\dagger}}_0+\eta$, with
$ \phi ^{\vphantom{\dagger}}_0=R_0(r)/r$.
This leads to the Bogoliubov equations \cite{sr96,bog2}
\begin{eqnarray}
\Omega\,u&=& \left ( - \frac{1}{2}\nabla^2+
\frac{1}{2} r^2-\mu-2| \phi ^{\vphantom{\dagger}}_0|^2
\right ) u-\phi_0^2(r)\, v
\nonumber\\
-\Omega\,v&=& \left ( - \frac{1}{2}\nabla^2+
\frac{1}{2} r^2-\mu-2| \phi ^{\vphantom{\dagger}}_0|^2 \right )
v-{\phi_0^*}^2(r)\, u
\label{bog}
\end{eqnarray}
where
\begin{equation}
\eta( {\mib r},t)=u( {\mib r})\,\exp(-i\Omega t)
+ v^*( {\mib r})\,\exp(i\Omega^* t)\ .
\end{equation}
The condensate function $ \phi ^{\vphantom{\dagger}}_0$
is taken to be real.  Note that
$(u,v)=( \phi ^{\vphantom{\dagger}}_0,- \phi^{\vphantom{\dagger}}_0)$
is a zero mode.

Taking advantage of the radial symmetry of $\phi^{\vphantom{\dagger}}_0)$,
one can classify the excited states by angular momentum and compute the
excitation energies; the spectrum was investigated in \cite{sr96}.
In figure \ref{fig6} we plot the three lowest values of $\Omega^2$
{\it versus\/} $\mu$.  Consistent with previous work \cite{sr96}, we 
find that for $\mu < 0.4$ there is a purely
imaginary eigenfrequency reflecting the instability of the stationary state.
The inverse of the imaginary part of $\Omega$ gives the typical time for the
instability to grow.  For larger values of the chemical potential (lower
values of $N$) the unstable mode crosses the zero mode and the spectrum
becomes that of a stable condensate, as in the repulsive case.  We find the
stability boundary lies at $\nu=0.57$, in agreement with the results of
ref. \cite{rhb95}.

\begin{figure}[!h]
\centering
\leavevmode
\epsfxsize=8cm
\epsfysize=8cm
\epsfbox[18 144 592 718] {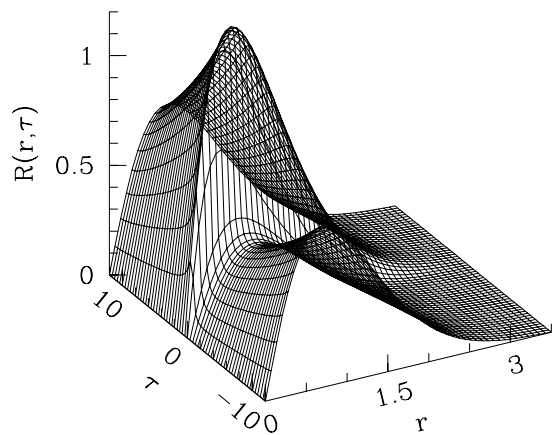}
\caption[]
{\label{fig7} Plot of the radial wave function of the bounce
for $\mu=0.7$. The system size
was $\tau \in [-12,12]$ and $r \in [0,3.5]$. The parameter $\delta$ (see
equation \ref{eqn:delta}) is 0.028.}
\end{figure}

We now compute the tunneling rate using the familiar instanton
formalism \cite{col85}.  In order to compute the instanton action,
we must solve the following equations of motion derived from equation
\ref{SE}.  Adding a chemical potential term, we obtain the equations
\begin{eqnarray}
- {\partial}_\tau\phi&=&- \frac{1}{2}\nabla^2\phi+
( \frac{1}{2} r^2-\mu)\,\phi- ( {\bar\phi}\phi)\phi
\label{eom1}\\
+ {\partial}_\tau {\bar\phi}&=&- \frac{1}{2}\nabla^2
{\bar\phi}+( \frac{1}{2} r^2-\mu)\, {\bar\phi}- ( {\bar\phi}\phi) {\bar\phi}
\label{eom2}
\end{eqnarray}
subject to periodic boundary conditions
on the interval $\tau\in[- \frac{1}{2}\beta,
\frac{1}{2}\beta]$, were $\beta$ is the
inverse temperature (we are interested in the $\beta\to\infty$ limit).
Note that these equations yield the imaginary time version of the continuity
equation,
\begin{equation}
i {\partial}_\tau\rho + \nabla\cdot  {\mib j}=0\ ,
\end{equation}
with
\begin{equation}
\rho= {\bar\phi}\phi \hskip 0.2cm, \hskip 0.2cm
{\mib j}={1\over 2i}( {\bar\phi}\nabla\phi-\phi\nabla {\bar\phi})\ .
\end{equation}
Thus, the total particle number $N=\int\!d^3\!r\,\rho( {\mib r})$ is conserved
by the evolution equations, since $ {\mib j}$ vanishes as $r\to\infty$.
It should be emphasized that equations \ref{eom1} and \ref{eom2} are
{\it not\/} complex conjugate pairs.  However, the instanton solutions
of interest all have the symmetry
\begin{equation}
 {\bar\phi}=\phi^*( {\mib r},-\tau)\ ,
\end{equation}
where $({}^*)$ denotes complex conjugation \cite{foot2}.
Substituting this into equation
\ref{eom1} leads to a nonlocal equation in (imaginary) time for $\phi$.
Nevertheless, a straightforward application of Newton's method may be used
to obtain a numerical solution \cite{fal97}.
\begin{figure}[!t]
\centering
\leavevmode
\epsfxsize=8cm
\epsfysize=8cm
\epsfbox[18 144 592 718] {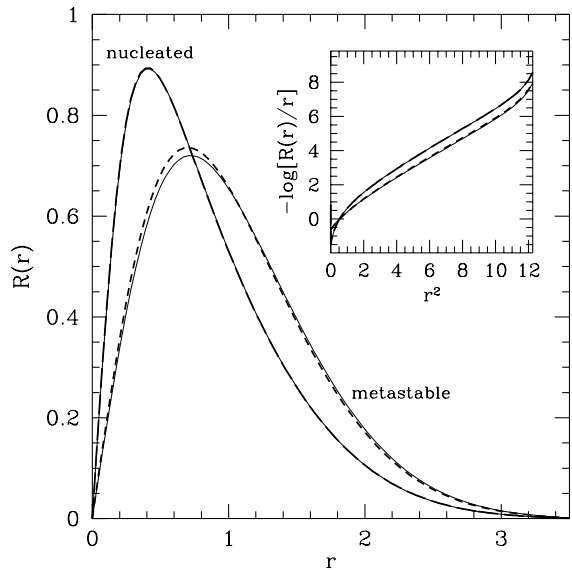}
\caption[]
{\label{fig8} The metastable state and the nucleated state for the
bounce of figure \ref{fig7}.  The inset shows $-\ln \phi$ {\it vs.\/} $r^2$;
deviations from pure Gaussian behavior are apparent.  Thin solid lines
show the corresponding results for periodic solutions.}
\end{figure}
The equations admit a static solution,
$\phi( {\mib r},\tau)= \phi ^{\vphantom{\dagger}}_0(r)$,
describing the metastable state.  In the limit $\beta\to\infty$, the
nontrivial bounce solution satisfies
$ \phi_{\rm b}({\mib r},\tau)= \phi ^{\vphantom{\dagger}}_0(r)$
at $\tau=\pm \frac{1}{2}\beta$, while describing a denser nucleated condensate
at $\tau=0$.  The tunneling rate, obtained by summing over all multibounce
solutions \cite{col85}, is
\begin{equation}
\Gamma=A\exp(-\Delta S ^{\vphantom{\dagger}}_ {\rm E}/\hbar)
\end{equation}
where $A={\rm det}^{-1/2}[\delta^2 S^{\vphantom{\dagger}}_{\rm E}]$
is the fluctuation determinant
prefactor, and $\Delta S ^{\vphantom{\dagger}}_{\rm E}$
is the difference between the Euclidean actions
of the bounce and static solutions to (\ref{eom1},\ref{eom2}).  Since the
energy is preserved by the dynamics, we obtain
\begin{equation}
{\Delta S ^{\vphantom{\dagger}}_ {\rm E}\over N\hbar}=
{\int d\tau\!\int d^3\!r\,  {\bar\phi}_{\rm b} {\partial}_\tau \phi_{\rm b}
\over\int d^3\!r\,  {\bar\phi}_{\rm b} \phi_{\rm b}}\ .
\label{SEnum}
\end{equation}

The numerical method used to solve for $ \phi_{\rm b}({\mib r},\tau)$
is described in ref. \cite{fal97}.  Briefly,  we look for real,
radially symmetric solutions, $ \phi_{\rm b}(r)=R_{\rm b}(r)/r$.
We discretize space-time and write
(\ref{eom1},\ref{eom2}) as a set of nonlinear equations which we solve
using the Newton-Powell method \cite{powell}.  This process demands an initial
guess for the bounce solution, which we took to be an interpolation between
the static $ \phi^{\vphantom{\dagger}}_0(r)$ at
$\tau=-\frac{1}{2}\beta$ and a somewhat denser
Gaussian profile at $\tau=0$.  It is necessary to choose this guess
to be asymmetric in imaginary time about $\tau=0$.

\begin{figure}[!t]
\centering
\leavevmode
\epsfxsize=8cm
\epsfysize=8cm
\epsfbox[18 144 592 718] {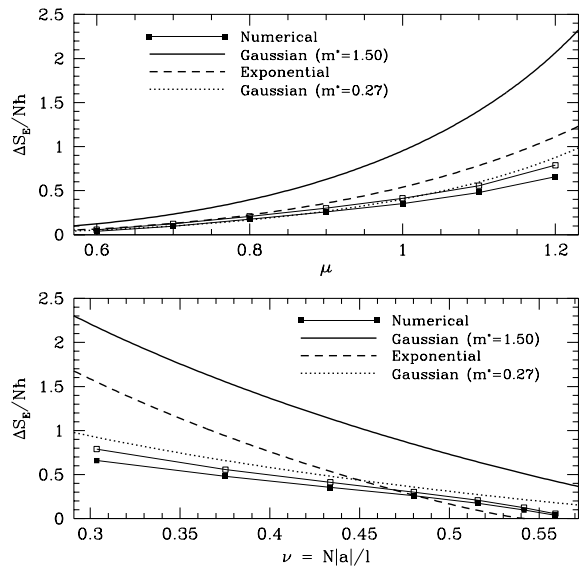}
\caption[]
{\label{fig9} A comparison of the numerically obtained instanton action
with various trial state predictions of
$\Delta S ^{\vphantom{\dagger}}_ {\rm E}/N\hbar$.
Filled squares represent numerical data with aperiodic boundary conditions;
open squares denote periodic boundary conditions.}
\end{figure}

When $\beta$ is finite, periodic solutions will deviate from the static
$\phi ^{\vphantom{\dagger}}_0$ at $\tau=\pm \frac{1}{2}\beta$.
We enforced $\phi(r,-\frac{1}{2}\beta)= \phi^{\vphantom{\dagger}}_0(r)$
which results in non-periodicity of the bounce solution.  We confirm that
as we increase $\beta$ that $\phi(r,+\frac{1}{2}\beta)$
approaches $ \phi^{\vphantom{\dagger}}_0(r)$ while
the central structure around $\tau\approx 0$ does not change appreciably.
It is the central part of the bounce which contributes most to
(\ref{SEnum}).  A measure of the aperiodicity is
\begin{equation}\label{eqn:delta}
\delta = {\int\limits_0^\infty dr\,|R_ {\rm b}(r,-\frac{1}{2}\beta)
- R_{\rm b}(r,+\frac{1}{2}\beta)|\over
\int\limits_0^\infty dr\,R_{\rm b}(r,-\frac{1}{2}\beta)} .
\end{equation}
We increased $\beta$ until $\delta< 0.05$, at which point we deemed our
solution a good approximation of the $\beta=\infty$ bounce.
The radial coordinate was chosen to extend from $r=0$ to a maximum
$r=L$, where $L$ was chosen to accommodate the metastable state.  We required
$R_{\rm b}(L,\tau) < 0.01$ in all cases, where typically we took $L=5$.

Another approach is to require periodicity for all $\beta$, as is necessary
when computing the finite temperature instanton solution.  In this case,
the wavefunctions at $\tau=\pm \frac{1}{2}\beta$ agree with each other,
but not with
the metastable solution of the NLSE.  Of course agreement is recovered in
the $\beta\to\infty$ limit.  In figures \ref{fig8} and \ref{fig9} we
compare the results of these two approaches for $\beta=24$.
We found the aperiodic solution systematically yielded a lower value of
$\Delta S ^{\vphantom{\dagger}}_ {\rm E}/N\hbar$.  The agreement between
the two schemes for $\beta=24$ is at the 30\% level, with better
agreement for larger values of $\mu$.

\begin{figure}[!h]
\centering
\leavevmode
\epsfxsize=8cm
\epsfysize=8cm
\epsfbox[18 144 592 718] {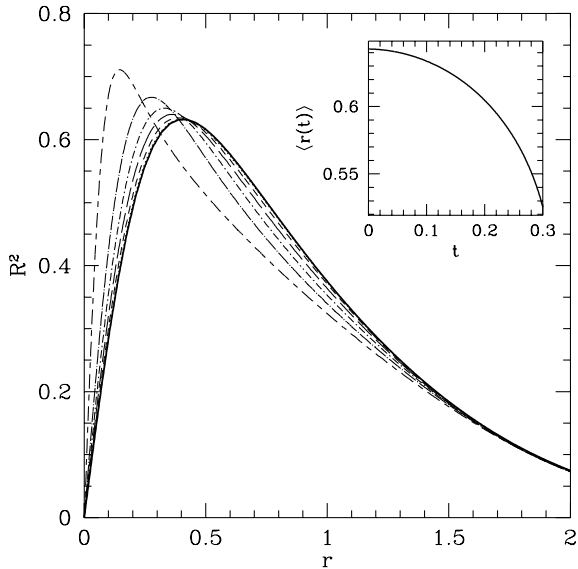}
\caption[]
{\label{fig10} Real time evolution of the nucleated condensate (dark solid
curve) for $t=0.05$, $t=0.10$, $t=0.15$, $t=0.20$, $t=0.25$, and $t=0.30$.
$N$ is accurately conserved by the integration.  The inset shows the collapse
of the condensate as measured by $\langle r(t)\rangle$.}
\end{figure}

In figure \ref{fig7} we plotted the radial wave function of the bounce,
$R_{\rm b}=r\, \phi_{\rm b}$, for $\mu=0.7$.
We stress that $ \phi_{\rm b}$
is interpretable as a condensate wavefunction only at
$\tau=\pm \frac{1}{2}\beta$
and at $\tau=0$, where it represents the metastable and nucleated condensates,
respectively.  In figure \ref{fig8} we plot, also for $\mu=0.7$,
the radial wavefunctions $R_0(r,- \frac{1}{2}\beta)$
and $R_ {\rm b}(r,0)$ corresponding
to the metastable and nucleated states, respectively.  Real time evolution
of the nucleated state using the nonlinear Schr{\"o}dinger equation describes
an ever collapsing condensate.  This is analogous to a tunneling particle
rolling down the hill after it emerges from the tunnel barrier.
In figure \ref{fig10} we show the radial condensate density, $r^2\,|\phi(r)|^2$,
developing towards collapse as the nucleated state evolves in real time
according to the NLSE \cite{foot3}.

We worked in the range $0.6 \le \mu \le 1.2$.  We could not reach larger
values of $\mu$ because these demanded a very large $\beta$ which increased
the calculation time \cite{foot4}.
For the smaller values of $\mu$ the bounce becomes flatter and loses its
temporal structure.  Eventually
$\Delta S^{\vphantom{\dagger}}_{\rm E}/N\hbar\to 0$ as $\mu$ decreases
to its minimum allowed value of $\mu^{\vphantom{\dagger}}_{\rm min}\simeq 0.4$.
In this region,
we only obtained convergence to the trivial (metastable) solution.

A comparison of our numerical results with those of the various collective
coordinate theories is shown in figure \ref{fig9}.  Apparently Stoof's
approximation $m^*=0.27$ is in much better agreement with the data than
$m^*=\frac{3}{2}$ derived from the Gaussian trial states of equations
(\ref{tri1}, \ref{tri2}).  No other parameters have been adjusted.

\section{Conclusions}

In this paper, we investigated the decay of metastable Bose-Einstein
condensates with attractive interactions confined in a parabolic trap.
The decay rate was calculated using the instanton formalism, applied to
the Gross-Pitaevskii action,
$S ^{\vphantom{\dagger}}_ {\rm E}[\psi, {\bar\psi}]$.  The field theory can
be reduced to an effective quantum mechanics problem by focusing on
a collective coordinate which describes the width of the condensate,
as was first done by Stoof \cite{sto97}.  We went beyond the collective
coordinate approach, numerically solving for the instanton field configuration
$\psi( {\mib r},\tau)$.  This was accomplished by rendering the nonlinear
partial differential equations for $\psi$ and $ {\bar\psi}$ as a large set of
nonlinear equations on a space-time lattice and solving them via the
Newton-Powell method.  We can thereby derive the nucleated state wavefunction,
which describes the condensate at the instant it emerges from the tunneling
barrier.  The nucleated state, evolved according to the real time nonlinear
Schr{\"o}dinger equation, collapses to an arbitrarily dense state for which
the attractive GP model is no longer a good approximation.  While the
numerically obtained metastable and nucleated states show deviations from
pure Gaussian behavior, a comparison of the leading contribution to the
semiclassical tunneling rate reveals moderate agreement between numerical
values and those obtained from Gaussian approximations to the (dynamical)
condensate wavefunction.  Adjusting only the collective coordinate
effective mass $m^*$ puts the two schemes in quite good agreement.
Thus we conclude that the collective coordinate description adequately
captures the essential physics of the tunneling process.

\section{Acknowledgements}
This work is in large part a continuation of work performed with H. Levine,
to whom we are grateful for numerous discussions and suggestions.  We also
thank D. S. Rokhsar and H. T. C. Stoof for discussions.
JAF acknowledges support from the Brazilian Agency CNPq.


\begin{references}
\bibitem{fn1} Of course, at atomic length scales the hard sphere nature
of the atoms is manifest and the potential is repulsive.
\bibitem{foot1} As $r\to 0$ there can be no linear term in $\psi(r)$,
so exponential behavior is wrong in the vicinity of the origin.
\bibitem{foot2} Because the metastable state was real we looked only for
real solutions.
\bibitem{foot3} The integration of the time dependent NLSE, equation
\ref{tdnlse}, was done using a split-step method. The Laplacian part is
integrated using a Crank-Nicholson method and the local part in
integrated exactly.  This results in scheme accurate to ${\cal O}(\Delta t)$.
\bibitem{foot4} The largest grid used had 30,000 points, 100 in the
radial direction and 300 in the time direction. All calculations were
performed in a PC running a Pentium 133MHz processor.
\end{references}
\end{document}